\documentstyle[prl,aps]{revtex}
\draft
\input epsf
\input{epsf.sty}
\begin{document}

\twocolumn[\hsize\textwidth\columnwidth\hsize\csname
@twocolumnfalse\endcsname

\title{
No Ending Point in The Bragg-to-Vortex Glass Phase Transition Line at Low Temperatures
}

\author{
S. L. Li, H. H. Wen\cite{responce}, Z. W. Zhao
}

\address{
National Laboratory for Superconductivity, Institute of Physics and Center for Condensed Matter Physics, Chinese Academy of Sciences, P.O. Box 603, Beijing 100080, China
}

\maketitle

\begin{abstract}
We have measured the magnetic hysteresis loops and the magnetic relaxation for $Bi_2Sr_2CaCu_2O_{8+\delta}$ (Bi-2212) single crystals which exhibit the second magnetization peak effect. Although no second peak effect is observed below 20 K in the measurement with fast field sweeping rate, it is found that the second peak effect will appear again after long time relaxation or in a measurement with very slow field sweeping rate at 16 K. It is anticipated that the peak effect will appear at very low temperatures ( approaching zero K ) when the relaxation time is long enough. We attribute this phenomenon to the profile of the interior magnetic field and conclude that the phase transition line of Bragg glass to vortex glass has no ending point at low temperatures.

\end{abstract}


]
The second peak ( SP ) effect in Bi-2212, which represents an anomalous increase of critical currents with the increase of the magnetic field, has attracted great interest because of its theoretical perplexity and practical prospect\cite{Chikumoto,Yang,Cubitt,Yeshurun}. Giamarchi {\it et al.}\cite{Giamarchi} have given a picture, in which the SP effect is explained as a consequence of phase transition between the low field Bragg glass and the high field vortex glass because the critical current of vortex glass is much greater than that of the Bragg glass. This theory interprets very well the experimental observation by Cubitt {\it et al.} that there is a quasi-ordered flux line lattice at low fields and low temperatures as seen by small-angle neutron diffraction\cite{Cubitt}. The measurements of local magnetization by using tiny Hall sensors also indicate a phase transition in the SP region\cite{Khaykovich}. Some recent experiments and discoveries, such as the dynamical observation of the SP effect by using magneto-optical imaging technique\cite{Beek}, the abrupt change of Josephson plasma frequency at the phase boundary of the Bragg glass\cite{Gaifullin}, the sample size effect of SP \cite{Wang}, etc., further support the picture based on the crossover from a Bragg glass to a vortex glass. However, despite the consistency between the theory and many experiments and the common belief that a phase transition occurs in the SP region, it remains unexplained why the SP disappears below about 20 K in the measurement of magnetic hysteresis loops ( MHL ) \cite{Goffman}. It seems that there exists an ending point at about 20 K in the Bragg-to-vortex glass phase transition line. This phenomenon has been related to a crossover to zero dimensional pinning, that is, the pinning of individual pancakes\cite{Goffman,Correa,Niderost} in low temperature region. In this scenario, there would be no SP effect at low temperatures at all. In this Letter we present strong evidence to show that the SP effect does exist at low temperatures. The only need to see the SP effect in low temperature region is to wait a long time to let vortex system comes close to the equilibrium state.

The Bi-2212 single crystals used in this paper were grown by the self-flux technique. The transitions are around 88 K with a transition width of 1 K. The samples have typical sizes of 3 mm $\times$ 3 mm and thickness of 20 $\mu m$. The magnetic measurements were carried out by a vibrating sample magnetometer ( VSM 8T, Oxford 3001 ) and a superconducting quantum interfere device ( SQUID, Quantum design, MPMS5.5 ). During the measurement the magnetic field is applied parallel to the c-axis of the single crystals. Fig. 1 shows the MHLs between 20 K and 40 K measured with a field sweeping rate of 100 Oe/s. It is clear that the SP effect appears at all temperatures above 20 K. There is no SP effect when $T < 20K$ in the general MHL measurement at all. 

To study the character of vortex matter at low temperatures, we measured the magnetic relaxation at several magnetic fields at T = 16 K, 18 K, 20 K and 14 K. It is well known that the Bean-Livingston surface barrier and geometrical barrier have much strong effect on the behavior of vortex matter in the ascending field branch than in the descending field branch of the MHL \cite{Zeldov}, therefore all the relaxation measurements were taken in the desce-
\begin{figure}[h]
	\vspace{10pt}
    \centerline{\epsfxsize 8cm \epsfbox{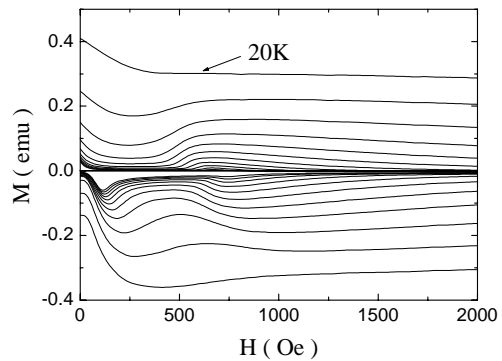}}
    \vspace{10pt}
\caption{
The MHL measured between 20 K and 40 K with a magnetic field sweeping rate of 100 Oe / s. No SP effect can be observed at T = 20 K.
}
\end{figure}
\begin{figure}[h]
	\vspace{10pt}
    \centerline{\epsfxsize 8cm \epsfbox{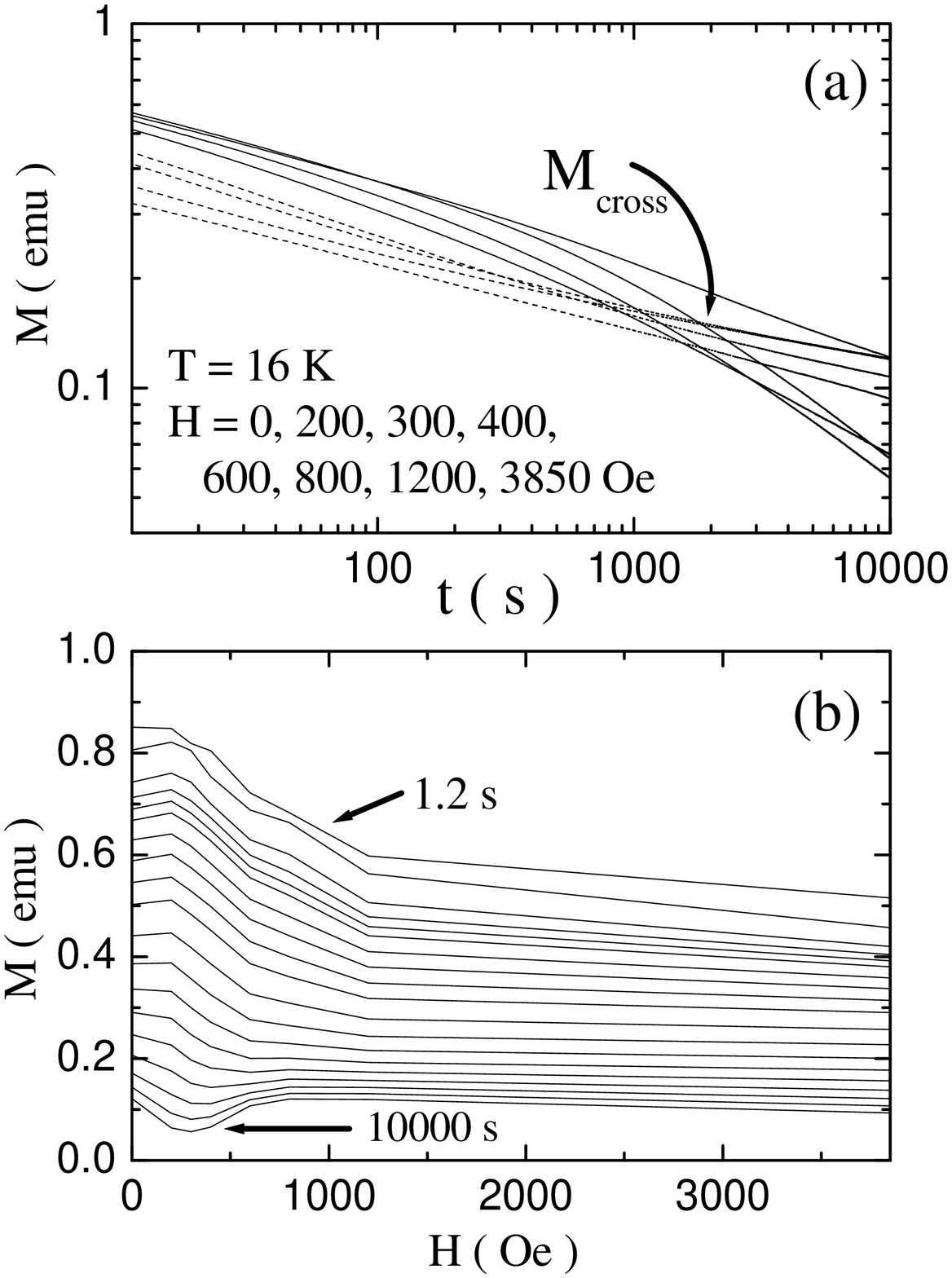}}
    \vspace{10pt}
\caption{
(a) The time dependence of magnetization measured at T = 16 K. The external fields ( from top to bottom at 10 s ) are 0, 200, 300, 400 Oe, plotted as solid lines and 600, 800, 1200, 3850 Oe, plotted as dotted lines respectively. The magnetization of low fields (solid lines) are larger than that of high fields (dotted lines) at a short time. After long time relaxation, the magnetization of low fields become smaller than that of high fields, and the SP effect appears. $M_{cross}$ represents the region where the crossover takes place. (b) The MHL transformed from the relaxation data. The SP effect is very clear at 16 K at a long relaxation time.
}
\end{figure}
\noindent nding field branch. That is, the field was first raised to 8 Tesla and then lowered to the expected field. The relaxation was then measured after the above process and the results are shown in Fig. 2(a). It is interesting that the magnetization at a lower field will gradually become smaller than that at a higher field after long time relaxation, which is manifested by a crossing area labeled as M$_{cross}$ in Fig. 2(a). It should be noted here that M$_{cross}$ represents only a region but not a point. This phenomenon clearly indicates that the SP effect will appear again if one waits a long time or the field sweeping rate in measuring the MHL is slow enough. Fig. 2(b) shows the MHL transformed from the data of Fig. 2(a), that is, we take the values of magnetization measured at the same time at different fields and plot them in the M-H coordinates. The emergence of the SP in Fig. 2(b) strongly suggests that the SP effect does exist at low temperatures. We have been able to observe the SP effect 
\begin{figure}[h]
	\vspace{10pt}
    \centerline{\epsfxsize 8cm \epsfbox{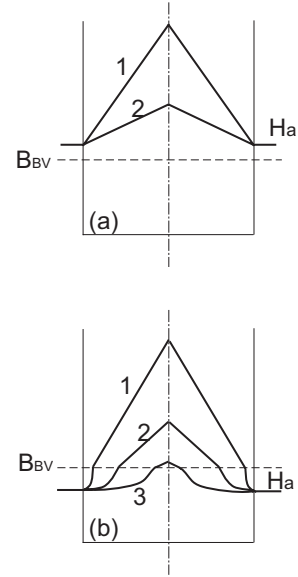}}
    \vspace{10pt}
\caption{
A schematic show for the interior magnetic field profiles during the relaxation. The dash line represents the phase transition field $B_{BV}$, and $H_a$ is the external field. The number 1, 2, 3 in the graph represent time flow, that is, $t_1 < t_2 < t_3$. (a) and (b) describe the conditions that $H_a > B_{BV}$ and $H_a < B_{BV}$ respectively.
}
\end{figure}
\noindent in the long time relaxation measurement at 16 K, 18 K and 20 K. At 14 K, however, the same tendency has been observed but no SP effect can be found since the relaxation time ( 10000s ) is not long enough. It is thus anticipated that the SP effect will appear when the magnetic relaxation is enough. In fact, Yeshurun {\it et al.}\cite{Yeshurun} have observed the similar phenomenon, which they explained as a dynamical character of the SP effect. It should be noted here that the $M_{cross}$ at three temperatures, 16 K, 18 K and 20 K are all approximately equal to 0.15 emu, which we will give an explanation below.

As we know, the vortex matter of Bi-2212 in the field sweeping process is far from the equilibrium state due to the bulk pinning. A commonly believed picture is the Bean critical state model, which is approximately applicable in the vortex glass state as observed recently by magneto-optical technique\cite{Beek}. Therefore, we give a sketch in Fig.3 to discuss the vortex dynamics based on Bean critical state model. In Fig.3, $B_{BV}$ is the hypothetical field at which Bragg glass changes to vortex glass. In our present sample $B_{BV}$ is a little less than 700 Oe. At  
\begin{figure}[h]
	\vspace{10pt}
    \centerline{\epsfxsize 8cm \epsfbox{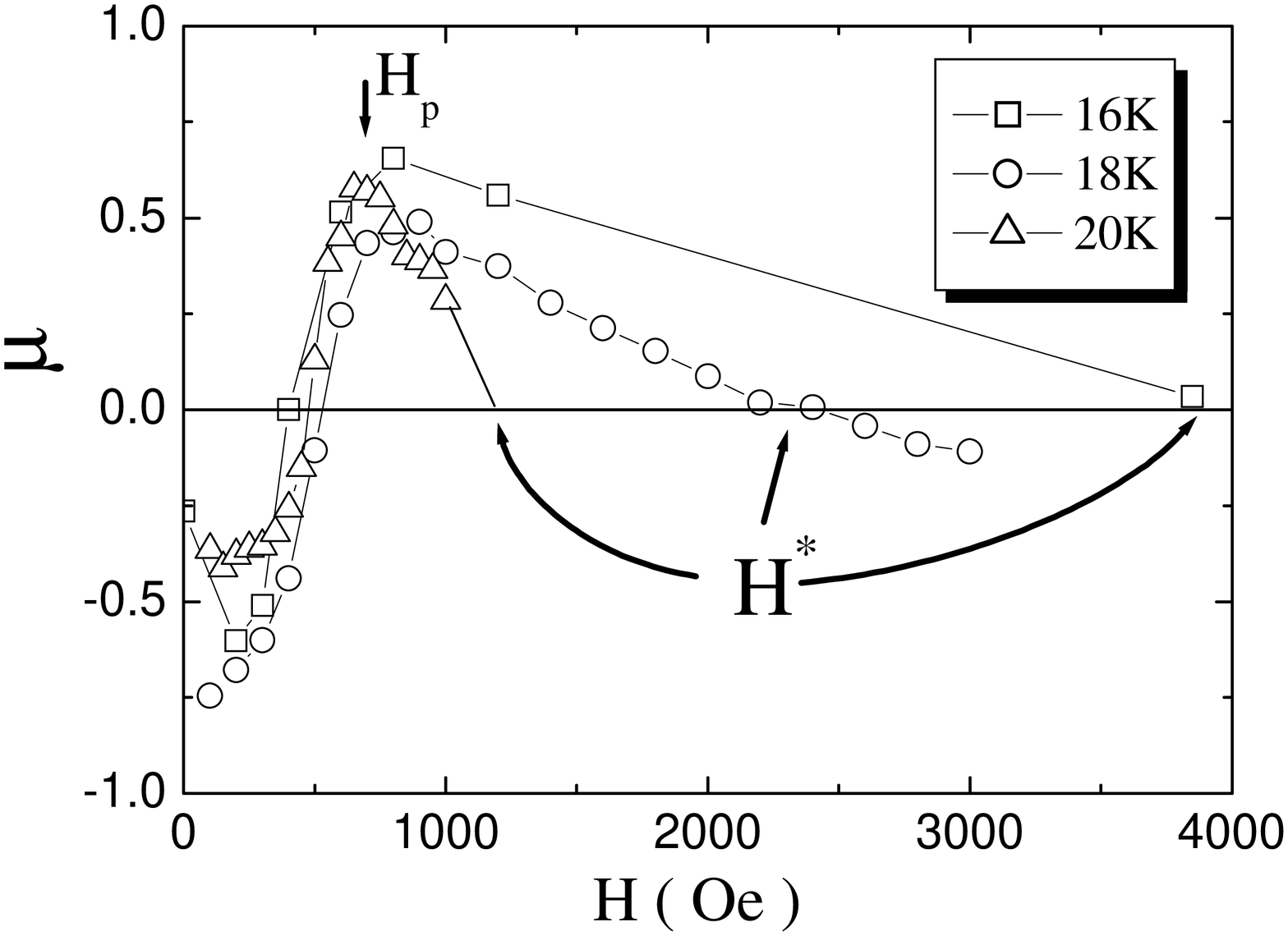}}
    \vspace{10pt}
\caption{
The characteristic exponent $\mu$ calculated by the Eq. (1) at 16, 18 and 20K. There are peaks on these curves, whose fields are equal to the peak field of SP, $H_p$. $H^*$ represents the crossover at which $\mu$ goes to zero.
}
\end{figure}
\noindent low temperatures, the flux creep rate is very slow and the gradient ( proportional to the superconducting current ) of the profile of interior magnetic field is very large. At high fields ($H_a > B_{BV}$), the whole vortex matter is in the vortex glass state,so the relaxation of the total magnetic moment, or equivalently the changing rate of the field profile are very slow ( as shown by dotted lines in Fig. 2(a) ). The interior field profiles are schematically shown by the curves labeled 1 and 2 in Fig. 3(a). But when $H_a < B_{BV}$, due to the large difference between the maximum field ( center ) and minimum field ( near edge ) within the sample, the vortex matter is partly in Bragg glass state and partly in the vortex glass state. In the very beginning of the relaxation process, the major part is in the vortex glass state, as shown by the curve labeled 1 in Fig. 3(b), therefore, the magnetic moment at a low applied field is larger than that at a high field and the relaxation is slow. During the relaxation, the part of Bragg glass will gradually grow up and finally become dominant, and the vortex matter will mainly be in the Bragg glass state at last, as the curve labeled 3 shown in Fig. 3(b). This explains why the magnetic moment measured at a low field decays faster and faster in the long time relaxation ( shown by the solid lines in Fig. 2(b)).  Because the critical current of Bragg glass is much less than that of vortex glass\cite{Giamarchi}, the SP effect will appear again. According to this picture, we think that the SP  effect will exist at very low temperatures, even T = 0 K. It is unfortunately not easy to check this issue directly in the magnetic and / or transport measurement. At a low temperature the waiting time required to see the SP effect will be far beyond the measurable time. 

This picture can also interpret the almost same $M_{cross}$ at three different temperatures. It is known that the transition field from the Bragg glass to the vortex glass is weakly temperature dependent. Therefore whatever the external applied field is, the only thing to determine 
\begin{figure}[h]
	\vspace{10pt}
    \centerline{\epsfxsize 8cm \epsfbox{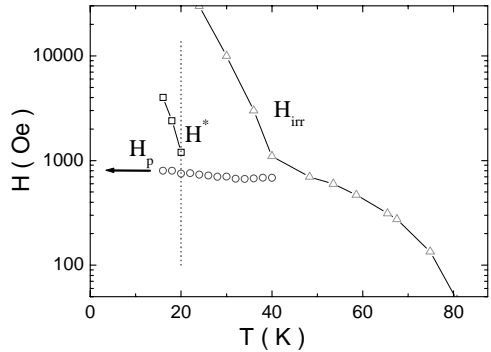}}
    \vspace{10pt}
\caption{
The vortex phase diagram of Bi-2212. $H_p$ is the SP field. From our result it is assumed that the SP effect will end up at 0 K, as indicated by the arrow in the graph. $H^*$ represents the field where $\mu$ = 0 and vortex system changes from the vortex glass state to another state ( probably the liquid2 state). 
}
\end{figure}
\noindent the appearance of the SP effect is the field profile. The crossing area near $M_{cross}$ may reflect the starting point of the dominance of the Bragg glass. According to this argument, we can estimate how long we have to wait for the appearance of the SP effect. We have measured the magnetic relaxation at 8 K and 300 Oe to 30000 s. If following the double-logarithmic decaying relation $ ln M = ln M_0 - A ln t$, it is found that we have to wait at least $10^14 s$ to see that the magnetic moment drops to 0.15 emu where the SP appears. It should be noted that above consideration is based on the simple Bean critical state model, the real relaxation process and profile of interior magnetic field are more complicated. 

In fact, some groups have already observed the phenomenon of co-existence of a quasi-ordered state and a disordered state and the existence of a phase boundary between them\cite{Giller,Konczykowski,Giller2}, which is in consistent with our picture.

To further investigate the properties of the vortex matter, we have derived the characteristic exponent $\mu$ from the relaxation data. The magnetization during the relaxation process is expressed by an interpolation formula\cite{Thompson}
\begin{equation}
M(t)^{-\mu}=(M')^{-\mu}+Cln(t)
\label{eq1}
\end{equation}
which $M'$ and C can be treated as constant approximately. The characteristic exponent $\mu$ at 16, 18 and 20K are plotted together in Fig. 4. It is interesting that there also exists a peak on the $\mu(H)$ curve, which exactly corresponds to the peak field $H_p$ in MHL. Below $H_p$, the value of $\mu$ continuously changes from negative to positive. We attribute the so-called negative $\mu$ value to the co-existence of Bragg glass and vortex glass, i.e., it is not a true value corresponding to a single vortex phase. Above $H_p$, there only exists vortex glass phase, the $\mu$ value was predicted to be $ 0 < \mu < 1$ \cite{VG}. It is interesting to note that, after a cusp at round 700 Oe, $\mu$ continuously drops down to zero and becomes negative at $H^*$. The reason of this zero and negative $\mu$ is unclear yet and is assumed to be a change of vortex dynamics\cite{slli}. This may have no direct relationship with SP effect.

Finally in Fig. 5 we give the vortex phase diagram based on the above analyze. The arrow represents the extrapolation of $H_p$ to very low temperatures. The dotted line shows the critical temperature below which the SP effect cannot be observed in the usual magnetization measurement. The irreversibility line  $H_{irr}$ is taken from Ref. \onlinecite{Shibauchi}.  The line $H^*$ shows the boundary between a positive $\mu$ value ( vortex glass ) and a negative one ( probably the liquid2 phase in high field region\cite{bouquet} ).  

In conclusion, the long time relaxation measurement clearly demonstrate that the SP effect can be observed in low temperature region ( down to 16 K ). The tendency found from the magnetization relaxation at even lower temperatures ( 14 K ) tells us that the SP effect will appear at very low temperatures if the vortex system is really close to the equilibrium state. Based on this explanation, we conclude that there is no ending point in the Bragg-to-vortex glass phase transition line at low temperatures. Our result here may further indicate that the disappearance of the SP effect in the usual magnetization measurement in low temperature region is not due to the zero-dimensional or pancake pinning, rather it is due to the large gradient of the magnetic induction within the sample.

\acknowledgements

This work is supported by the National Science Foundation of China (NSFC 19825111) and the Ministry of Science and Technology of China ( project: NKBRSF-G1999064602 ). HHW acknowledges fruitful discussion with Dr. T. Giamarchi.

\end{document}